\documentclass{article}
\usepackage{graphicx} 
\usepackage{xcolor}%
\usepackage{amsthm}
\usepackage{amsmath}
\usepackage{amssymb}
\usepackage{thmtools}
\usepackage{thm-restate}
\usepackage{hyperref}

\newtheorem{theorem}{Theorem}
\newtheorem{lemma}[theorem]{Lemma}

\usepackage[
    backend=biber,
    style=alphabetic,
    maxbibnames=99,
    maxalphanames=4
]{biblatex}
\title{Closing Open Shop up to constant factors}
\author{Mikhail Cherniavskii \qquad Mikhail Koshelev \qquad
Andrey Kupavskii \\ Stanislav Moiseev \qquad
Fedor Ozhegov \qquad Alexander Shekhovtsov}
\date{}

\begin{document}

\maketitle
\begin{abstract}
    We resolve, up to constant factors, a longstanding question on sufficient load conditions for solving the nonpreemptive open-shop scheduling problem exactly in polynomial time with makespan equal to the maximum machine load. 
    
    Let $m$ be the number of machines, $M$ the maximum machine load, and $K$ the longest operation. The classical algorithm of Fiala (1983) provides a sufficient bound $M \geq \Omega(Km\log m)$. We prove that $M\ge23Km$ suffices for constructing an optimal schedule of makespan $M$ in deterministic polynomial  time. This matches, up to a constant factor, the lower bound
$2K(m-1)$ of Sevastyanov (1995).
\end{abstract}

\section{Introduction}
We study nonpreemptive open-shop makespan minimization, denoted by
$O\|C_{\max}$ in the three-field notation of Graham, Lawler, Lenstra, and Rinnooy Kan~\cite{GrahamEtAl1979}.
An instance consists of $n$ jobs and $m$ machines. Job~$j$ has one
operation of processing time $p_{ij}\geq 0$ on each machine~$i$.
The operations of a job may be performed in any order (hence the name \textit{open} shop). Each operation
is \emph{nonpreemptive}: once it starts, it runs to completion.
At any time, each machine processes at most one job and each job uses
at most one machine. Let $C_j$ denote the completion time of the last
operation of job~$j$. The objective is to minimize the \emph{makespan}
$C_{\max}=\max_j C_j$. When the number $m$ of machines is fixed, we
write $O_m\|C_{\max}$.

Let $M=\max_i\sum_j p_{ij}$ be the maximum machine load,
$J=\max_j\sum_i p_{ij}$ the maximum job load, and
$K=\max_{i,j}p_{ij}$ the longest operation. Every schedule has
makespan at least $\max\{M,J\}$. Gonzalez and Sahni~\cite{GonzalezSahni1976}
showed that this bound is attainable in polynomial time for the
preemptive problem $O\mid\mathrm{pmtn}\mid C_{\max}$ and in linear
time for $O_2\|C_{\max}$. Already $O_3\|C_{\max}$ is NP-hard.
Since $J\leq mK$, the preemptive optimum equals $M$ whenever
$M\geq mK$. We study the load sufficient to attain $M$ when operations
must run uninterrupted.

\paragraph{Main result.}
We prove that $M\geq 23mK$ guarantees a nonpreemptive schedule of
makespan $M$ (Theorem~\ref{th: main}). 
For $m\geq 3$, let $\eta(m)$ be the least $\rho$ such that every
$m$-machine instance with $M\geq \rho K$ has optimal makespan $M$.
The question of how small the sufficient load multiplier can be goes
back to B\'ar\'any and Fiala~\cite[p.~180]{BaranyFiala1982}.
Woeginger~\cite[Open Problem~6]{Woeginger2018} later posed the problem
of obtaining sharp upper and lower bounds on $\eta(m)$.
Together with the lower bound of Sevastyanov~\cite{Sevastyanov1995},
our theorem gives
\[
  2m-2 \leq \eta(m) \leq 23m,
\]
determining the asymptotic order of the threshold. We do not attempt to optimize the constant $23$ in our bound.
With the assistance of GPT-6 Pro, we also obtained the stronger
upper bound $\eta(m)\leq 6m$. However, its proof is substantially
longer, and we omit it in favor of a simpler and more transparent
argument establishing the asymptotically tight bound above.

\paragraph{Previous load bounds.}
For each value of $\rho$
below, the corresponding algorithm constructs a schedule of makespan
$M$ whenever $M\geq\rho K$.
\begin{center}
\renewcommand{\arraystretch}{1.15}
\begin{tabular}{@{}ll@{}}
\hline
Algorithm & Sufficient load multiplier $\rho$ \\
\hline
Fiala~\cite{Fiala1983} & $2^{\lceil\log_2 m\rceil}(16\lceil\log_2 m\rceil+5)$ \\
B\'{a}r\'{a}ny and Fiala~\cite{BaranyFiala1982} & $2^{\lceil\log_2 m\rceil}(8\lceil\log_2 m\rceil+5)$ \\
Sevastyanov~\cite{Sevastyanov1992} &
  $\frac{16}{3}m\log_2 m+\frac{61}{9}m-7.4$ \\
Sevastyanov~\cite{Sevastyanov1995} & $m^2-1+\frac{1}{m-1}$ \\
This work & $23m$ \\
\hline
\end{tabular}
\end{center}
The 1992 bound is given in the form recorded by Kononov, Sevastyanov,
and Tchernykh~\cite{KononovEtAl1999}. The quadratic bound
is useful for small $m$. For example, it gives $\eta(3)\leq 17/2$,
and a separate three-machine algorithm attains $M$ in $O(n\log n)$
time when $M\geq 7K$~\cite{Sevastianov1998Nonstrict}. As a function
of $m$, the earlier results gave $\eta(m)=O(m\log m)$, leaving a
logarithmic gap above the lower bound $2m-2$. Ahmadian, Khatami, Salehipour, and Cheng~\cite{AhmadianEtAl2021} survey these load conditions
alongside other tractable cases of the open shop.


\paragraph{Related algorithms.}
Racsm\'{a}ny's greedy algorithm gives a $2$-approximation for
$O\|C_{\max}$~\cite{BaranyFiala1982,Woeginger2018}. If operation $(i,j)$ finishes
last, every idle interval on machine $i$ before that operation starts
is covered by operations of job $j$ on other machines. The greedy
makespan is therefore at most $M+(m-1)K$. Williamson, Hall, Hoogeveen, Hurkens, Lenstra, Sevastyanov,
and Shmoys~\cite{WilliamsonEtAl1997} ruled out every polynomial-time
approximation ratio strictly below $5/4$ unless
$\mathrm{P}=\mathrm{NP}$. For each fixed $m$,
Sevastyanov and Woeginger~\cite{SevastyanovWoeginger1998} gave a
polynomial-time approximation scheme for $O_m\|C_{\max}$.
Dong, Jin, Lin, Su, Tong, and Xu~\cite{DongEtAl2025} gave an efficient polynomial-time
approximation scheme for identical open shops operating in parallel.
Each job is assigned to one shop, and both the number of shops and
the number of machines per shop are fixed.

For the \emph{proportionate} open shop, where $p_{ij}=p_j$ on every
machine, Sevastyanov~\cite{Sevastyanov2019} constructed a
schedule of makespan $M$ in linear time for fixed $m$ whenever
$M\geq(2m-2)K$, and proved that this multiplier is sharp for
every $m\geq3$. The proportionate case is easier because all machines
have the same operation lengths. A fixed order of the jobs therefore
has the same load in each of its initial segments on every machine,
so the machines are synchronized automatically. This feature does not
carry over to arbitrary processing times. When $p_{ij}$ depends on
$i$, the loads of an initial segment can differ between two machines
by an amount of order $M$, and keeping them balanced on all machines
at once is the main difficulty in the general case.

On three machines, Sevastyanov~\cite{Sevastyanov2019} solved the
proportionate problem $O_3\mid\mathrm{prpt}\mid C_{\max}$ in
pseudopolynomial time. Ni and Chen~\cite{NiChen2020} gave a
polynomial-time $13/12$-approximation for the same problem. For arbitrary
processing times on three machines, Kononov, Sevastianov, and Tchernykh~\cite{KononovEtAl1999} showed that a gap of at least $2K$
between the heaviest and lightest machine loads suffices to construct
a schedule of makespan $M$ in linear time. Drobouchevitch simplified the
linear-time exact algorithm for instances in which each job uses a
common machine and one of the other two~\cite{Drobouchevitch2021}. For a broader account
of complexity and approximation in open shops, see
Strusevich~\cite{Strusevich2022}.

\paragraph{Removing the logarithmic factor.}
Our construction follows the approach of B\'ar\'any and
Fiala~\cite{BaranyFiala1982} and Fiala~\cite{Fiala1983}. The jobs are
divided into two classes, and within each class the operations are
assigned to an ordered sequence of \emph{cells}. Every machine
processes the cells in order, alternating between the two classes, so
that the work of one class separates consecutive cells of the other.
For this to give a feasible schedule, the operations of each job must
lie in distinct cells, and on every machine the load of each initial
segment of cells must be close to its proportional share. The deviation
from this share is the error that has to be controlled.

The assignment to cells is built recursively: a range of cells is cut
into two parts, and a rounding step decides which operations go to
each part. In prior work every level of the recursion adds its own
rounding error, so the total error is proportional to the depth of
the recursion, which is of order $\log m$. This is the source of the
logarithmic factor in the earlier bounds. Our main contribution is a
recursive assignment whose error does not grow with the depth. At
every step we push the error in the right direction: the split on each
machine is chosen to compensate the error made so far instead of
adding to it. Hence the errors do not accumulate, and the final error
is a constant multiple of $K$ (Lemma~\ref{lem:cells}).

\paragraph{Organization.}
Section~\ref{sec:main} states the main theorem and proves it assuming
the cell assignment lemma (Lemma~\ref{lem:cells}).
Section~\ref{sec:rounding} proves the two rounding lemmas for bipartite
graphs required to prove Lemma~\ref{lem:cells}. Section~\ref{sec:cells} proves
Lemma~\ref{lem:cells}.

\section{Main Theorem}
\label{sec:main}

Consider a nonpreemptive open shop with $m$ machines and $n$ jobs.
Job $j$ consists of $m$ operations. Its $i$th operation has to be
processed on machine $i$ and takes time $p_{ij}\ge 0$.
We denote this operation by $(i,j)$. Let $H_i = \sum_{j=1}^n p_{ij}$ be
the total load of machine $i$, and set
\begin{align*}
    K &= \max_{i,j} p_{ij},
    &
    M &= \max_i H_i.
\end{align*}
Thus, $K$ is the duration of the longest operation and $M$ is the
largest machine load.

\begin{theorem}
\label{th: main}
If $M\ge 23mK$, then the open shop admits a schedule of makespan $M$.
Moreover, the schedule can be constructed in
deterministic polynomial time.
\end{theorem}

The proof is based on an assignment of operations to \emph{cells},
which we describe first.

\paragraph{Cells.}
Instead of choosing start times directly, we first attach to every
operation a label from $\{1,\dots,2m\}$, which we call its
\emph{cell}. A machine will process its operations in the increasing order of their assigned cells. 
If machine $i$ has total load $H_i$, then ideally every cell carries
the same load $h_i=H_i/(2m)$ on this machine, so that machine $i$
completes cells $1,\dots,u$ exactly at time $u \cdot h_i$.
We ask for two properties of the labelling.
\begin{itemize}
\item \emph{Balance.} On every machine $i$ and for every $u$, the
  load of cells $1,\dots,u$ differs from the ideal value $u \cdot h_i$ by
  at most a constant multiple of $K$, independently of $m$ and $n$.
\item \emph{Distinctness.} The operations of each job lie in pairwise
  distinct cells.
\end{itemize}

Lemma~\ref{lem:cells} supplies us with a labelling that satisfies two above properties. Note that if we try to use the schedule produced by this labelling, then we might have two operations of the same job processed at the same time. Even though we have a distinctness property, there might be some error on the joints between the cells and operations in two consecutive cells might collide with each other.

To overcome this issue, we do the trick used in \cite{Fiala1983}. We partition jobs into two classes and apply Lemma~\ref{lem:cells} to each class to get the labelling. We then interleave cells of different classes. More precisely, we construct a new labelling that places cells in the following order: the $1$st cell of the first class, the $1$st of the second class, the $2$nd cell of the first class, the $2$nd cell of the second class, and so on. Balance property guarantees us that a machine $i$ finishes processing operations assigned in the first $u$ cells at a time $u \cdot h_i + O(K)$. The cells $u$ and $u + 1$ contain operations that belong to different jobs due to interleaving, so an overlap may happen only for a job containing operations in cells $u$ and $u + 2$. However, because $h_i$ is chosen to be $\Omega(K)$, the collision may not take place.

We now state the lemma that labels operations.

\begin{restatable}[Cell assignment]{lemma}{cellslemma}
\label{lem:cells}
Let $p_{ij}\ge 0$ for $i\in[m]$, $j\in[n]$, and $K=\max_{i,j}p_{ij}$.
For each machine $i$ let
\[
H_i=\sum_{j=1}^{n}p_{ij}
\qquad\text{and}\qquad
h_i=\frac{H_i}{2m}
\]
be its total load and its target load per cell, and assume that
$h_i\ge \frac{11}{2}K$ for every~$i$.
Then there are injective maps $\sigma_j\colon[m]\to[2m]$, one for
each job $j$, such that $\sigma_j(i)$ is the cell assigned to operation $(i, j)$, with the following property. Let
\[
S_i(u)=\sum_{j\,:\,\sigma_j(i)\le u}p_{ij}
\]
be the load of machine $i$ in cells $1,\dots,u$. Then for every
$i\in[m]$ and every $u\in\{0,1,\dots,2m\}$,
\[
\bigl|S_i(u)-u \cdot h_i\bigr|\le E_u,
\qquad
E_u=\begin{cases}
K, & u \text{ even},\\[2pt]
\tfrac32 K, & u \text{ odd}.
\end{cases}
\]
\end{restatable}


We prove Lemma~\ref{lem:cells} in Section~\ref{sec:cells}. Here we use
it to prove Theorem~\ref{th: main}.

\begin{proof}[Proof of Theorem~\ref{th: main}]
Assume $K>0$. Otherwise, any assignment works. 

On each machine of load less than $M$, increase processing times until
the load equals $M$, keeping every processing time at most $K$.
This is possible because $M\le nK$. We write $p_{ij}$ for
the enlarged processing times. 

\medskip
\paragraph{Partition the jobs into two classes.}
Let $x_j\in[0,1]$ denote the fraction of job $j$ assigned to class
$J_0$, with the remaining fraction assigned to $J_1$. Consider the
polytope defined by
\begin{align*}
    \sum_{j=1}^{n} p_{ij}x_j &= \frac{M}{2}
        && \text{for every machine } i,\\
    0\le x_j &\le 1
        && \text{for every job } j.
\end{align*}
It is nonempty because $x_j=1/2$ for every $j$ is feasible.
Choose an extreme point $x$ of this polytope. At most $m$ coordinates
of $x$ are fractional because there are $m$ load constraints.


We round each $x_j$ to the nearest $y_j\in\{0,1\}$, breaking ties arbitrarily. Put job in $J_0$ if $y_j = 1$ and in $J_1$ otherwise.
Write $H_{i,c}$ for the total load of class $J_c$ on machine $i$.
Since at most $m$ coordinates change, each by at most $1/2$, we have
\begin{align*}
    \left|H_{i,0}-\frac{M}{2}\right|
    &= \left|\sum_{j=1}^{n}p_{ij}(y_j-x_j)\right|\\
    &\le \sum_{j:\,0<x_j<1}p_{ij}|y_j-x_j|
     \le \frac{mK}{2}.
\end{align*}
The same bound holds for $H_{i,1}$ because $H_{i,1}=M-H_{i,0}$.
Thus, for every machine $i$ and each $c\in\{0,1\}$, the average load
per cell satisfies
\begin{align*}
    h_{i,c}:=\frac{H_{i,c}}{2m}
    \ge \frac{M-mK}{4m}
    \ge \frac{11K}{2},
\end{align*}
where the last inequality follows from $M\ge 23mK$.

\medskip
\paragraph{Assigning operations to cells and constructing the schedule.}
We apply Lemma~\ref{lem:cells} separately to $J_0$ and $J_1$.
For each job $j\in J_c$, the lemma gives an injective map
$$
    \sigma_j^c:\{1,\ldots,m\}\longrightarrow\{1,\ldots,2m\},
$$
where $\sigma_j^c(i)$ is the cell assigned to the operation of job $j$
on machine $i$. Let
$$
    R_c^u=\bigl\{(i,j):j\in J_c,\ \sigma_j^c(i)=u\bigr\}.
$$
On every machine, we process the cells in the order $1,\ldots,2m$.
In cell $u$, we first run all operations of $R_0^u$ assigned to that
machine, then all operations of $R_1^u$ assigned to it.


\medskip
Let us show that the constructed schedule satisfies all requirements. First note that each machine works without idle time, so the maximum makespan is $M$. It remains to show that no two operations of the same job are being processed at the same time.

Let $S_i^c(u)$ denote the total processing time of the operations from $\bigcup_{w\le u} R^w_c$ on machine $i$, and set
$S_i^c(0)=0$. By Lemma~\ref{lem:cells},
\begin{align*}
    S_i^c(u) &= uh_{i,c}+e_i^c(u),
    &
    |e_i^c(u)| &\le E_u,
    &
    E_u &=
    \begin{cases}
        K,    & u \text{ even},\\
        3K/2, & u \text{ odd}.
    \end{cases}
\end{align*}
Fix a class $c\in\{0,1\}$, two machines $i,\ell$, and a cell $u$ with
$1\le u<2m$. We need to show that the operations of $R^u_c$ on machine $i$
finish no later than the operations of $R^{u+1}_c$ start on machine
$\ell$ (see Figure~\ref{fig:cell-separation}).

\begin{figure}[t]
    \centering
    \includegraphics[width=\linewidth]{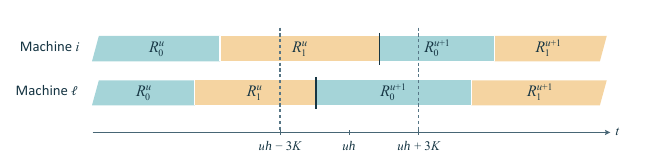}
    \caption{Transition from cell $u$ to cell $u+1$ on machines
    $i$ and $\ell$. Each solid tick marks the end of cell $u$ on its
    machine. Across all machines, the dashed line at $u\cdot h-3K$ separates
    $R_0^u$ from $R_0^{u+1}$, and the dashed line at $u\cdot h+3K$
    separates $R_1^u$ from $R_1^{u+1}$.}
    \label{fig:cell-separation}
\end{figure}

We first express this claim through the prefix loads. Every machine
works without idle time and, within each cell, runs the operations of
class $0$ before those of class $1$. Hence:
\begin{itemize}
\item on machine $i$, the operations of $R_0^u$ finish at time
  $S_i^0(u)+S_i^1(u-1)$, and the operations of $R_1^u$ finish at time
  $S_i^0(u)+S_i^1(u)$;
\item on machine $\ell$, the operations of $R_0^{u+1}$ start at time
  $S_\ell^0(u)+S_\ell^1(u)$, and the operations of $R_1^{u+1}$ start at
  time $S_\ell^0(u+1)+S_\ell^1(u)$.
\end{itemize}
So we have to prove the following two inequalities, the first for
class $0$ and the second for class $1$:
\begin{align}
    S_i^0(u)+S_i^1(u-1) &\le S_\ell^0(u)+S_\ell^1(u),
    \label{eq:sep0}\\
    S_i^0(u)+S_i^1(u) &\le S_\ell^0(u+1)+S_\ell^1(u).
    \label{eq:sep1}
\end{align}

To prove them, set $h=M/(2m)$, so that $h_{i,0}+h_{i,1}=h$ for every
machine~$i$, and substitute $S_i^c(u)=uh_{i,c}+e_i^c(u)$.

Let us expand the inequality \eqref{eq:sep0}. The two sides are
\begin{align*}
    S_i^0(u)+S_i^1(u-1)
    &= uh_{i,0}+(u-1)h_{i,1}+e_i^0(u)+e_i^1(u-1)\\
    &= uh-h_{i,1}+e_i^0(u)+e_i^1(u-1),\\
    S_\ell^0(u)+S_\ell^1(u)
    &= uh_{\ell,0}+uh_{\ell,1}+e_\ell^0(u)+e_\ell^1(u)\\
    &= uh+e_\ell^0(u)+e_\ell^1(u).
\end{align*}
Hence the right-hand side of~\eqref{eq:sep0} minus its left-hand side
equals
\begin{align*}
    h_{i,1}+e_\ell^0(u)+e_\ell^1(u)-e_i^0(u)-e_i^1(u-1)
    \ge h_{i,1}-3E_u-E_{u-1}.
\end{align*}

Now, we expand the inequality \eqref{eq:sep1}. The two sides are
\begin{align*}
    S_i^0(u)+S_i^1(u)
    &= uh+e_i^0(u)+e_i^1(u),\\
    S_\ell^0(u+1)+S_\ell^1(u)
    &= (u+1)h_{\ell,0}+uh_{\ell,1}+e_\ell^0(u+1)+e_\ell^1(u)\\
    &= uh+h_{\ell,0}+e_\ell^0(u+1)+e_\ell^1(u).
\end{align*}
Hence the right-hand side of~\eqref{eq:sep1} minus its left-hand side
equals
\begin{align*}
    h_{\ell,0}+e_\ell^0(u+1)+e_\ell^1(u)-e_i^0(u)-e_i^1(u)
    \ge h_{\ell,0}-3E_u-E_{u+1}.
\end{align*}

It remains to check that both lower bounds are nonnegative. Since $u$
has parity opposite to both $u-1$ and $u+1$, we have
$3E_u+E_{u\pm1}\le 3\cdot\frac32K+K=\frac{11}{2}K$. On the other hand,
$h_{i,1}\ge\frac{11}{2}K$ and $h_{\ell,0}\ge\frac{11}{2}K$.

\end{proof}

\section{Rounding Lemma}
\label{sec:rounding}

Our construction recursively splits the current set of operations
into two parts. To control the load discrepancies at each split,
we use the following rounding lemma for weighted bipartite graphs.
The lemma slightly generalizes Fiala's division lemma~\cite{Fiala1983}
and also follows from Knuth's two-way rounding lemma~\cite{Knuth1995}. An unweighted version is given by
Gandhi, Khuller, Parthasarathy, and Srinivasan~\cite{GandhiEtAl2006}. For background on iterative
rounding, we refer to the monograph by Lau, Ravi, and
Singh~\cite{LauRaviSingh2011}. For completeness, we include
a proof using path rounding in the spirit of
\cite{GandhiEtAl2006,srinivasan2008budgeted}.

We start with the special case of a more general lemma. For this case, we obtain an improved bound of $\frac12$ of maximum price.

\begin{lemma}[Halving, right degree-$2$]
\label{lem:halve}
Let $G=(U\cup W,E)$ be a finite bipartite multigraph in which every
vertex of $W$ has degree at most two, and let $p_e\ge 0$ for $e\in E$.
Then there is $y\in\{0,1\}^E$ such that the two edges at every vertex
of $W$ of degree two receive different values, and for every $u\in U$
\[
\Bigl|\sum_{e\sim u}p_e\bigl(y_e-\tfrac12\bigr)\Bigr|
\le\frac12\max_{e\sim u}p_e ,
\]
where the right-hand side is zero if $u$ is isolated.
\end{lemma}

\begin{proof}
For each $u\in U$ list its edges as $e_1,\dots,e_s$ with
$p_{e_1}\ge\dots\ge p_{e_s}$ and group them into the consecutive pairs
$\{e_1,e_2\},\{e_3,e_4\},\dots$. If $s$ is odd, $e_s$ stays unpaired.
Let $\Gamma$ be the multigraph with vertex set $E$ in which two edges
of $G$ are joined if they form such a pair at a vertex of $U$, and
also if they are the two edges at a vertex of $W$ of degree two. 
Every $e\in E$ is in at most one pair of each kind, so $\Gamma$ has
maximum degree at most two, and along every cycle of $\Gamma$ the two
kinds of pairs alternate. Hence all cycles of $\Gamma$ are even and
$\Gamma$ has a proper colouring $y\colon E\to\{0,1\}$.

The pairs at $W$ give the first claim. For the second, fix $u\in U$
and abbreviate $p_t=p_{e_t}$, with $p_{s+1}=0$. The two edges of a
pair $\{e_{2t-1},e_{2t}\}$ receive different values, so together they
contribute $\pm\frac12(p_{2t-1}-p_{2t})$ to the sum. An unpaired
edge $e_s$ contributes $\pm\frac12p_s$. Therefore
\[
\Bigl|\sum_{e\sim u}p_e\bigl(y_e-\tfrac12\bigr)\Bigr| \le \frac{1}{2}\sum\limits_{t\ge 1} |p_{2t-1} - p_{2t}|
=\frac12\Bigl(p_1-\sum_{t\ge1}(p_{2t}-p_{2t+1})\Bigr)
\le\frac12\,p_1 ,
\]
where the sums use $p_t=0$ for $t>s$ and the that the
$p_t$ are nonincreasing. 
\end{proof}

Now, we prove the general case.

\begin{lemma}[Halving, arbitrary degree]
\label{lem:split}
Let $G=(U\cup W,E)$ be a finite bipartite multi-graph, and let
$f(e,v)\ge0$ be the load of edge $e$ at its endpoint $v$.
Every fractional assignment $x\in[0,1]^E$ can be rounded to
$y\in\{0,1\}^E$ so that
\[
\left|\sum_{e\sim v}f(e,v)(y_e-x_e)\right|
\leq \max_{e\sim v}f(e,v)
\]

\end{lemma}

\begin{proof}
Initially, set $z \leftarrow x$. We iteratively round $z$ to a point in $\{0, 1\}^E$. Define $\operatorname{frac}(z) \subseteq E$ to be the set of edges $e$, such that $0 < z_e < 1$. Define the discrepancy of a vector $z \in \mathbb{R}^E$:
\[
\Delta_v(h)=\sum_{e\sim v}f(e,v)\cdot h_e.
\]
Suppose there is a non-zero vector $h \in \mathbb{R}^E$, such that $\Delta_v(h) = 0$ for all $v \in U \cup W$ and $\operatorname{supp}(h) \subseteq \operatorname{frac}(z)$. Then, we shift the current $z$ by the vector $h$ until some new variable gets rounded: $z \leftarrow z + h \cdot t$. Suppose there is no anymore a vector that we can slide along. Let us see that the subgraph $(U \cup W, \operatorname{frac}(z))$ contains at most one cycle per component. Assume there is a component $C$ for which the number of edges, $|E(C)|$, is more than the number of vertices, $|V(C)|$. Then, there is a non-zero vector $h \in \mathbb{R}^E$ with $\operatorname{supp}(h) \subseteq E(C)$, such that $\Delta_v(h) = 0$ for all $v \in V(C)$. Moreover, $\Delta_v(h) = 0$ holds for all $v \in U \cup W$ because other vertices are not incident to edges in $C$. So, each component of the subgraph contains at most one cycle. Since the graph is bipartite, these cycles are of even length. For each cycle $C$, let $e_1, \ldots, e_{2l}$ be its edges in a consecutive order. We round variables of the cycle alternately: $z_{e_1} \leftarrow 0$, $z_{e_2} \leftarrow 1$, $z_{e_3} \leftarrow 0$, $z_{e_4} \leftarrow 1$, $\ldots$, $z_{2l - 1} \leftarrow 0, z_{2l} \leftarrow 1$. To this end, note that for a vertex $v$, the error $|\Delta_v(z - x)|$ is at most $\max_{e \sim v} f(e, v)$, and the subgraph $(U \cup W, \operatorname{frac}(z))$ is a forest with at most one vertex with non-zero error in a component. (See Figure \ref{fig:cycle-forest-path}).

\begin{figure}[htbp]
    \centering
    \includegraphics[width=\linewidth]{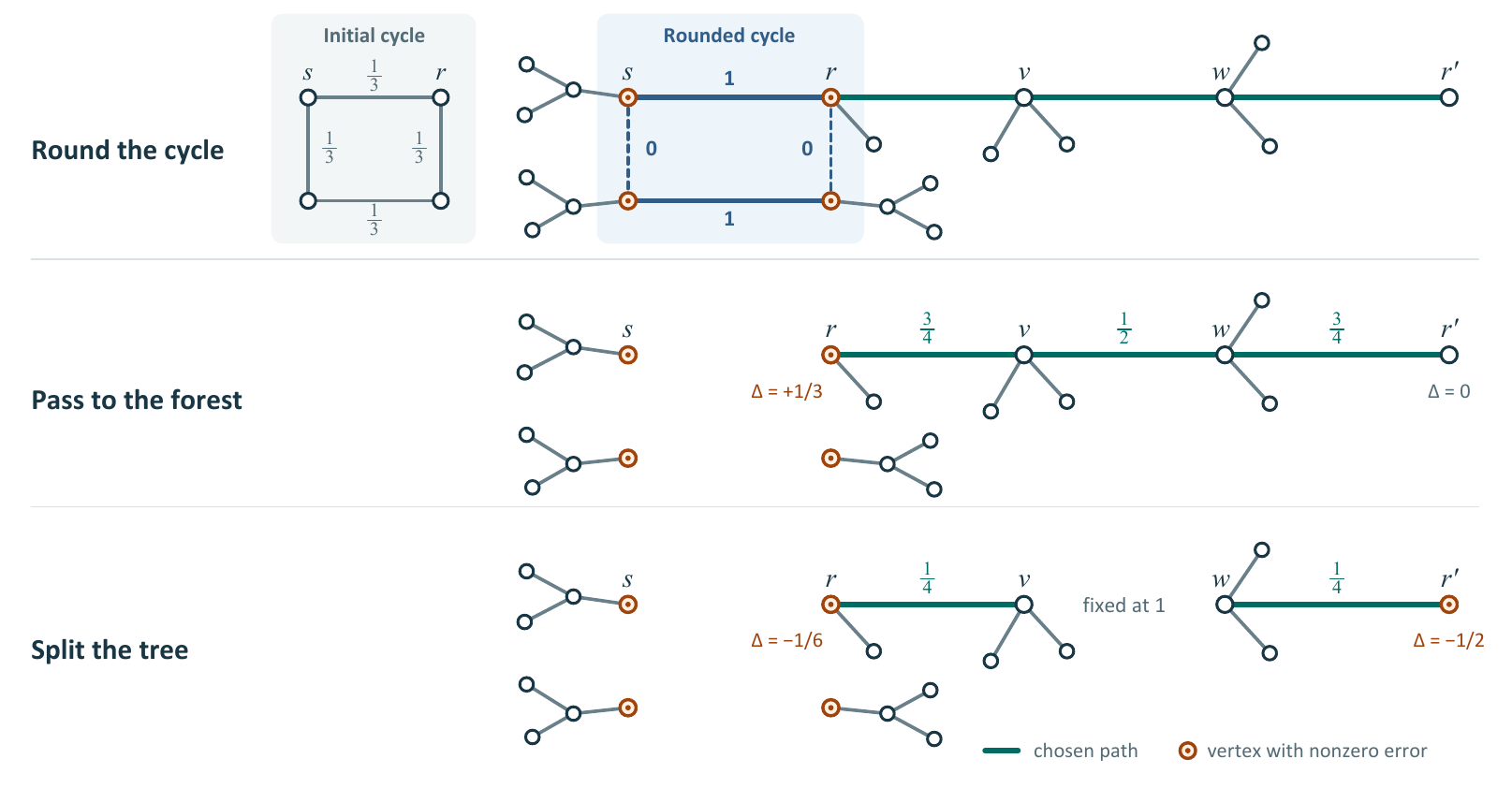}
    \caption[Cycle rounding and path splitting]{%
        Cycle rounding and path splitting in the proof of Lemma~\ref{lem:split}.
        Initially, each cycle edge has value $1/3$, the path edges
        $rv$, $vw$, and $wr'$ have values $3/4$, $1/2$, and $3/4$,
        respectively. Every edge outside the cycle and this path has value $1/2$.
        All endpoint weights are $1$ except $f(sr,s)=2$.
        Rounding the cycle edges alternately to $1$ and $0$ leaves a
        fractional forest with one vertex of nonzero error in each tree.
        The path update fixes $vw$ at $1$. Removing this edge from the
        fractional forest places $r$ and $r'$ in separate trees. 
    }
    \label{fig:cycle-forest-path}
\end{figure}

We maintain these as an invariant: the subgraph $(U \cup W, \operatorname{frac}(z))$ is a
forest, each component contains at most one vertex with nonzero
error, and every such vertex $v$ satisfies $|\Delta_v(z - x)| \leq \max_{e \sim v} f(e, v)$.
In a component containing an edge, let $r$ be its vertex of
nonzero error, if any, and otherwise any leaf.
Choose a path $P$ from $r$ to a distinct leaf $r'$.
Take a nonzero $h\in\mathbb{R}^E$, supported on $P$, satisfying
\[
\Delta_v(h) = 0
\qquad(v\in V(P)\setminus\{r,r'\}).
\]
Such a direction exists since the number of linear equations ($|V(P)| - 2$) is less than the number of variables $(|V(P)| - 1$). Without loss of generality we can assume that $\Delta_r(z - x) \cdot \Delta_r(h) \leq 0$. If it is not the case, then we can replace $h$ by $-h$. Let us slide along direction $h$ until some variable gets rounded: $z \leftarrow z + t \cdot h$ where $t > 0$. By assumption, the error can only be incurred in vertices $r$ and $r'$. The vertex $r'$ had no error before, so after the slide, its error $|\Delta_{r'}(z - x)|$ can be bounded by $\max_{e \sim r'} f(e, r')$. Since we have chosen $h$, so that $\Delta_r(h)$ has the opposite sign to $\Delta_r(z - x)$, the error for vertex $r$ can also be bounded by $\max_{e \sim r} f(e, r)$.

Let $e$ be the edge that was rounded: $z_e \in \{0, 1\}$. Note that the invariant is maintained: the tree has been cut into two and each of these trees contain at most one vertex with non-zero error bounded by $\max_{e \sim r} f(e, r)$ and $\max_{e \sim r'}f(e, r)$, respectively. We continue doing this procedure until every variable gets rounded. At the end we return the final $z$ as the integral vector $y$.

\end{proof}

\section{Recursive Division}
\label{sec:cells}

In this section we prove Lemma~\ref{lem:cells}, which we restate for
convenience.

\cellslemma*



\begin{proof}[Proof of Lemma~\ref{lem:cells}]
Assume $K>0$, as otherwise we can set $\sigma_j$ to be any injective maps. 

We construct the maps $\sigma_j$ recursively. At each step we are
given an interval of cells $(a,b]=\{a+1,\dots,b\}$ with
even endpoints $a<b$, together with the set
$R\subseteq[m]\times[n]$ of operations that are to be placed in these
cells. Every operation outside $R$ has already been committed either
to $(0,a]$ or to $(b,2m]$. Hence the prefix loads $S_i(a)$ and
$S_i(b)$ are already determined, no matter how the operations of $R$
are later distributed inside $(a,b]$.

We maintain two invariants:
\begin{itemize}
\item[(I1)] $|S_i(a)-a \cdot h_i|\le K$ and $|S_i(b)-b \cdot h_i|\le K$ for
  every machine $i$;
\item[(I2)] every job has at most $(b-a)/2+1$ operations in $R$.
\end{itemize}

We start with $(a,b]=(0,2m]$ and $R=[m]\times[n]$. Here $S_i(0)=0$,
$S_i(2m)=H_i=2m \cdot h_i$, and every job has $m=(b-a)/2$ operations, so
both invariants hold. Let us describe the step of the recursion.

\paragraph{Case $b-a\ge4$.} In this case we define a cutting point: let
$k_0=2\lfloor(b-a)/4\rfloor$, $k_1=b-a-k_0$, and $c=a+k_0$. We divide
$R=R_0\sqcup R_1$ and recurse on $(a,c]$ with $R_0$ and on $(c,b]$
with $R_1$. Let us explain how we make the division
$R=R_0\sqcup R_1$.

Our aim is to get 
$S_i(c) = c \cdot h_i \pm K$ for every machine $i$. The operations of $R$
on machine $i$ have total load $S_i(b)-S_i(a)$. If exactly
$c \cdot h_i-S_i(a)$ of this load went to $R_0$, we would get
$S_i(c)=c \cdot h_i$. So, ideally, $R_0$ should receive the fraction
\[
\lambda_i=\frac{c \cdot h_i-S_i(a)}{S_i(b)-S_i(a)}
\]
of the load of machine $i$.

We round the fractional vector $(\lambda_i)_{i=1}^m$ into an integer one. We apply
Lemma~\ref{lem:split} to the bipartite graph whose vertices are the
machines and the jobs and whose edges are the operations in $R$. The
edge $(i,j)$ has load $p_{ij}$ at machine $i$, load $1$ at job $j$,
and fractional value $x_{(i,j)}=\lambda_i$. We let $R_0$ consist of
the operations rounded to $1$, and $R_1$ of those rounded to $0$ (see Figure \ref{fig:cell-assignment-rounding}).
\begin{figure}[htbp]
    \centering
    \includegraphics[width=\linewidth]{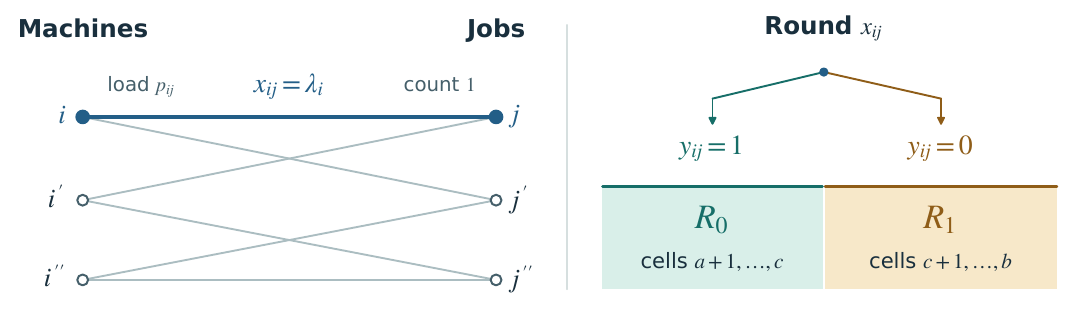}
    \caption{Rounding an operation at a recursive cut.
    The edge $(i,j)$ has weight $p_{ij}$ at machine $i$ and weight $1$
    at job $j$. Its fractional value $x_{ij}=\lambda_i$ is rounded to
    $y_{ij}\in\{0,1\}$. If $y_{ij}=1$, the operation enters $R_0$
    and is assigned to a cell in $(a,c]$. If $y_{ij}=0$, it enters
    $R_1$ and is assigned to a cell in $(c,b]$.}
    \label{fig:cell-assignment-rounding}
\end{figure}

The lemma gives two guarantees.
\begin{itemize}
\item[(a)] For every machine $i$, the load of $R_0$ on machine $i$
  differs from $c \cdot h_i-S_i(a)$ by at most $K$.
\item[(b)] For every job, the number of its operations in $R_0$
  differs by less than $1$ from the sum of the fractions $\lambda_i$
  of its operations in $R$. The same holds for $R_1$ with the
  fractions $1-\lambda_i$.
\end{itemize}
To apply the lemma we need $\lambda_i\in[0,1]$. Below we also need
that $\lambda_i$ is close to $k_0/(b-a)$, the fraction of cells that
lie in $(a,c]$. We prove both facts at the end of this step:
\begin{equation}
\label{eq:lambda}
0<\lambda_i<1
\qquad\text{and}\qquad
\Bigl|\lambda_i-\frac{k_0}{b-a}\Bigr|\le\frac{1}{5(b-a)} .
\end{equation}

We now check that the invariants hold again for $(a,c]$ with $R_0$
and for $(c,b]$ with $R_1$. Both $k_0$ and $k_1$ are even, so the new
intervals have even endpoints. The property (a) guarantees that the invariant \textup{(I1)} holds.

Let us see that the invariant \textup{(I2)} holds as well. Let $t\in\{0,1\}$. The interval that receives $R_t$ has $k_t$ cells,
so we have to show that every job has at most $k_t/2+1$ operations in
$R_t$.

An operation on machine $i$ goes to $R_0$ with fraction $\lambda_i$
and to $R_1$ with fraction $1-\lambda_i$. By~\eqref{eq:lambda}, in
both cases this fraction is at most
\[
\frac{k_t}{b-a}+\frac{1}{5(b-a)} .
\]
By (I2), a job has at most $(b-a)/2+1$ operations in $R$. Hence the
sum of the fractions of its operations is at most
\[
\Bigl(\frac{b-a}{2}+1\Bigr)
\Bigl(\frac{k_t}{b-a}+\frac{1}{5(b-a)}\Bigr)
=\frac{k_t}{2}+\frac{k_t}{b-a}+\frac{1}{10}+\frac{1}{5(b-a)}
<\frac{k_t}{2}+1 .
\]
The last inequality uses $k_t\le\frac23(b-a)$ and $b-a\ge4$, which
bound the three middle terms by $\frac23+\frac1{10}+\frac1{20}<1$.
By guarantee (b), the number of operations of the job in $R_t$
exceeds this sum by at most $1$. So it is an integer smaller than
$k_t/2+2$, that is, at most $k_t/2+1$.

Now we turn to prove \emph{\eqref{eq:lambda}.}
By (I1) and $h_i\ge\frac{11}{2}K$,
\[
c \cdot h_i-S_i(a)\ge k_0h_i-K>0
\qquad\text{and}\qquad
S_i(b)-c \cdot h_i\ge k_1h_i-K>0 .
\]
The first quantity is the numerator of $\lambda_i$, and the second is
the denominator minus the numerator. Hence $0<\lambda_i<1$.

For the second claim, note first that $k_0+k_1=b-a$ and
$c=a+k_0=b-k_1$, and therefore
\[
(b-a)\cdot c=k_1\cdot c+k_0\cdot c=k_1(a+k_0)+k_0(b-k_1)=k_1a+k_0b .
\]
We have
\begin{align*}
\lambda_i-\frac{k_0}{b-a}
&=\frac{c\cdot h_i-S_i(a)}{S_i(b)-S_i(a)}-\frac{k_0}{b-a}\\[4pt]
&=\frac{(b-a)\bigl(c\cdot h_i-S_i(a)\bigr)-k_0\bigl(S_i(b)-S_i(a)\bigr)}
       {(b-a)\bigl(S_i(b)-S_i(a)\bigr)}\\[4pt]
&=\frac{(b-a)\cdot c\cdot h_i-k_1S_i(a)-k_0S_i(b)}
       {(b-a)\bigl(S_i(b)-S_i(a)\bigr)}\\[4pt]
&=\frac{(k_1a+k_0b)\cdot h_i-k_1S_i(a)-k_0S_i(b)}
       {(b-a)\bigl(S_i(b)-S_i(a)\bigr)}\\[4pt]
&=-\,\frac{k_1\bigl(S_i(a)-a\cdot h_i\bigr)+k_0\bigl(S_i(b)-b\cdot h_i\bigr)}
          {(b-a)\bigl(S_i(b)-S_i(a)\bigr)} .
\end{align*}
In the third line we collected the terms with $S_i(a)$, using
$(b-a)-k_0=k_1$. In the fourth line we used the identity for
$(b-a)\cdot c$ above.

By (I1) and $k_0+k_1=b-a$, the numerator is at most $(b-a)K$ in
absolute value. In the denominator, again by (I1),
\[
S_i(b)-S_i(a)\;\ge\;(b-a)\,h_i-2K\;\ge\;\tfrac{11}{2}(b-a)K-2K\;\ge\;5(b-a)K ,
\]
where the last inequality uses $b-a\ge4$. This
proves~\eqref{eq:lambda} and completes the step for $b-a\ge4$.

\paragraph{Case $b-a=2$.} In this case, the interval consists of the two cells $a+1$ and $b$, and
we place every operation of $R$ into one of them. By (I2), every job
has at most two operations in $R$. We apply Lemma~\ref{lem:halve} to
the bipartite graph of machines and jobs with edge set $R$ and
weights $p_{ij}$. Operations with $y=1$ go to cell $a+1$, the others
to cell $b$.

The lemma guarantees, first, that two operations of the same job are
placed in different cells. Second, for every machine $i$ the load
placed in cell $a+1$ differs from half of its load in $R$ by at most
$K/2$, that is,
\[
\Bigl|S_i(a+1)-\frac{S_i(a)+S_i(b)}{2}\Bigr|\le\frac K2 .
\]
Since $a+1=(a+b)/2$, invariant (I1) gives
\[
\Bigl|\frac{S_i(a)+S_i(b)}{2}-(a+1)\,h_i\Bigr|\le K ,
\]
and therefore $|S_i(a+1)-(a+1)\,h_i|\le\frac32K$.



\end{proof}

\section*{Statement of AI usage}
ChatGPT 6 Pro was used to assist in developing the mathematical proofs and preparing the manuscript.

\printbibliography

@article{Sevastyanov2019,
  author  = {Sevastyanov, Sergey},
  title   = {Some Positive News on the Proportionate Open Shop Problem},
  journal = {Siberian Electronic Mathematical Reports},
  volume  = {16},
  pages   = {406--426},
  year    = {2019},
  doi     = {10.33048/semi.2019.16.023}
}

@article{NiChen2020,
  author  = {Ni, Guanqun and Chen, Lei},
  title   = {Improved Scheduling for the Three-Machine Proportionate Open Shop and Mixed Shop Minimum Makespan Problems},
  journal = {IEEE Access},
  volume  = {8},
  pages   = {186805--186812},
  year    = {2020},
  doi     = {10.1109/ACCESS.2020.3030694}
}

@article{Drobouchevitch2021,
  author  = {Drobouchevitch, Inna G.},
  title   = {Three-Machine Open Shop with a Bottleneck Machine Revisited},
  journal = {Journal of Scheduling},
  volume  = {24},
  number  = {2},
  pages   = {197--208},
  year    = {2021},
  doi     = {10.1007/s10951-020-00652-9}
}

@article{AhmadianEtAl2021,
  author  = {Ahmadian, Mohammad Mahdi and Khatami, Mostafa and Salehipour, Amir and Cheng, T. C. E.},
  title   = {Four Decades of Research on the Open-Shop Scheduling Problem to Minimize the Makespan},
  journal = {European Journal of Operational Research},
  volume  = {295},
  number  = {2},
  pages   = {399--426},
  year    = {2021},
  doi     = {10.1016/j.ejor.2021.03.026}
}

@article{Strusevich2022,
  author  = {Strusevich, Vitaly A.},
  title   = {Complexity and Approximation of Open Shop Scheduling to Minimize the Makespan: A Review of Models and Approaches},
  journal = {Computers \& Operations Research},
  volume  = {144},
  pages   = {105732},
  year    = {2022},
  doi     = {10.1016/j.cor.2022.105732}
}

@article{DongEtAl2025,
  author  = {Dong, Jianming and Jin, Ruyan and Lin, Guohui and Su, Bing and Tong, Weitian and Xu, Yao},
  title   = {An Efficient Polynomial-Time Approximation Scheme for Parallel Multi-Stage Open Shops},
  journal = {Discrete Applied Mathematics},
  volume  = {377},
  pages   = {390--401},
  year    = {2025},
  doi     = {10.1016/j.dam.2025.08.004}
}

@article{GonzalezSahni1976,
  author  = {Gonzalez, Teofilo and Sahni, Sartaj},
  title   = {Open Shop Scheduling to Minimize Finish Time},
  journal = {Journal of the ACM},
  volume  = {23},
  number  = {4},
  pages   = {665--679},
  year    = {1976},
  doi     = {10.1145/321978.321985}
}

@article{BaranyFiala1982,
  author  = {B{\'a}r{\'a}ny, Imre and Fiala, Tibor},
  title   = {T{\"o}bbg{\'e}pes {\"u}temez{\'e}si probl{\'e}m{\'a}k k{\"o}zel optim{\'a}lis megold{\'a}sa},
  journal = {Szigma},
  volume  = {15},
  number  = {3},
  pages   = {177--191},
  year    = {1982},
  note    = {In Hungarian},
  url     = {https://journals.lib.pte.hu/index.php/szigma/article/view/2716}
}

@article{Sevastyanov1992,
  author  = {Sevastyanov, S. V.},
  title   = {A Polynomial-Time Open-Shop Problem with an Arbitrary Number of Machines},
  journal = {Cybernetics and Systems Analysis},
  volume  = {28},
  number  = {6},
  pages   = {918--933},
  year    = {1992},
  doi     = {10.1007/BF01291296}
}

@article{Sevastyanov1995,
  author  = {Sevastyanov, Sergey},
  title   = {Vector Summation in {Banach} Space and Polynomial Algorithms for Flow Shops and Open Shops},
  journal = {Mathematics of Operations Research},
  volume  = {20},
  number  = {1},
  pages   = {90--103},
  year    = {1995},
  doi     = {10.1287/moor.20.1.90}
}

@inproceedings{Woeginger2018,
  author    = {Woeginger, Gerhard J.},
  title     = {The Open Shop Scheduling Problem},
  booktitle = {35th Symposium on Theoretical Aspects of Computer Science ({STACS} 2018)},
  editor    = {Niedermeier, Rolf and Vall{\'e}e, Brigitte},
  series    = {Leibniz International Proceedings in Informatics},
  volume    = {96},
  pages     = {4:1--4:12},
  publisher = {Schloss Dagstuhl -- Leibniz-Zentrum f{\"u}r Informatik},
  year      = {2018},
  doi       = {10.4230/LIPIcs.STACS.2018.4}
}

@article{GrahamEtAl1979,
  author  = {Graham, R. L. and Lawler, E. L. and Lenstra, J. K. and {Rinnooy Kan}, A. H. G.},
  title   = {Optimization and Approximation in Deterministic Sequencing and Scheduling: A Survey},
  journal = {Annals of Discrete Mathematics},
  volume  = {5},
  pages   = {287--326},
  year    = {1979},
  doi     = {10.1016/S0167-5060(08)70356-X}
}

@article{WilliamsonEtAl1997,
  author  = {Williamson, D. P. and Hall, L. A. and Hoogeveen, J. A. and Hurkens, C. A. J. and Lenstra, J. K. and Sevastyanov, S. V. and Shmoys, D. B.},
  title   = {Short Shop Schedules},
  journal = {Operations Research},
  volume  = {45},
  number  = {2},
  pages   = {288--294},
  year    = {1997},
  doi     = {10.1287/opre.45.2.288}
}

@article{SevastyanovWoeginger1998,
  author  = {Sevastyanov, Sergey V. and Woeginger, Gerhard J.},
  title   = {Makespan Minimization in Open Shops: A Polynomial Time Approximation Scheme},
  journal = {Mathematical Programming},
  volume  = {82},
  number  = {1-2},
  pages   = {191--198},
  year    = {1998},
  doi     = {10.1007/BF01585871}
}

@article{Sevastianov1998Nonstrict,
  author  = {Sevastyanov, Sergey},
  title   = {Nonstrict Vector Summation in Multi-Operation Scheduling},
  journal = {Annals of Operations Research},
  volume  = {83},
  pages   = {179--212},
  year    = {1998},
  doi     = {10.1023/A:1018908013582}
}

@article{KononovEtAl1999,
  author  = {Kononov, A. and Sevastyanov, S. and Tchernykh, I.},
  title   = {When Difference in Machine Loads Leads to Efficient Scheduling in Open Shops},
  journal = {Annals of Operations Research},
  volume  = {92},
  pages   = {211--239},
  year    = {1999},
  doi     = {10.1023/A:1018986731638}
}

@inproceedings{srinivasan2008budgeted,
  title={Budgeted allocations in the full-information setting},
  author={Srinivasan, Aravind},
  booktitle={International Workshop on Approximation Algorithms for Combinatorial Optimization},
  pages={247--253},
  year={2008},
  organization={Springer}
}

@article{Knuth1995,
  author  = {Knuth, Donald E.},
  title   = {Two-Way Rounding},
  journal = {SIAM Journal on Discrete Mathematics},
  year    = {1995},
  volume  = {8},
  number  = {2},
  pages   = {281--290},
  doi     = {10.1137/S0895480194264757},
  url     = {https://arxiv.org/abs/math/9504228}
}

@article{Fiala1983,
  author  = {Fiala, Tibor},
  title   = {An Algorithm for the Open-Shop Problem},
  journal = {Mathematics of Operations Research},
  year    = {1983},
  volume  = {8},
  number  = {1},
  pages   = {100--109},
  doi     = {10.1287/moor.8.1.100},
  url     = {https://pubsonline.informs.org/doi/10.1287/moor.8.1.100}
}

@article{GandhiEtAl2006,
  author  = {Gandhi, Rajiv and Khuller, Samir and Parthasarathy, Srinivasan and Srinivasan, Aravind},
  title   = {Dependent Rounding and Its Applications to Approximation Algorithms},
  journal = {Journal of the ACM},
  year    = {2006},
  volume  = {53},
  number  = {3},
  pages   = {324--360},
  doi     = {10.1145/1147954.1147956},
  url     = {https://www.cs.umd.edu/~srin/PDF/2005/depround-dec05.pdf}
}

@book{LauRaviSingh2011,
  author    = {Lau, Lap Chi and Ravi, R. and Singh, Mohit},
  title     = {Iterative Methods in Combinatorial Optimization},
  publisher = {Cambridge University Press},
  year      = {2011},
  url       = {https://cs.uwaterloo.ca/~lapchi/papers/iterative.pdf}
}

\end{document}